\documentclass[a4paper]{spie}  %>>> use this instead for A4 paper
\usepackage{amsmath,amsfonts,amssymb}
\usepackage{graphicx}
\usepackage{multirow}
\usepackage[colorlinks=true, allcolors=blue]{hyperref}
\usepackage{lineno}
\usepackage{xcolor}
\usepackage{setspace}
\newcommand\farcm{\mbox{$.\mkern-4mu^\prime$}}% 
\let\farcm\farcm 
\newcommand\farcs{\mbox{$.\!\!^{\prime\prime}$}}% 
\let\farcs\farcs 
\newcommand\degc{$^\circ$C}

  {}             % 終了時の設定

\title{Initial operations of the Soft X-ray Imager onboard XRISM}

\author[a]{Hiromasa Suzuki}
\author[b]{Tomokage Yoneyama}
\author[c]{Shogo B. Kobayashi}
\author[d]{Hirofumi Noda}
\author[e]{Hiroyuki Uchida}
\author[f]{Kumiko K. Nobukawa}
\author[g]{Kouichi Hagino}
\author[h]{Koji Mori}
\author[a]{Hiroshi Tomida}
\author[i]{Hiroshi Nakajima}
\author[j]{Takaaki Tanaka}
\author[k]{Hiroshi Murakami}
\author[l]{Hideki Uchiyama}
\author[m]{Masayoshi Nobukawa}
\author[a]{Yoshiaki Kanemaru}
\author[h]{Yoshinori Otsuka}
\author[h]{Haruhiko Yokosu}
\author[h]{Wakana Yonemaru}
\author[h]{Hanako Nakano}
\author[h]{Kazuhiro Ichikawa}
\author[h]{Reo Takemoto}
\author[h]{Tsukasa Matsushima}
\author[n]{Marina Yoshimoto}
\author[n]{Mio Aoyagi}
\author[n]{Kohei Shima}
\author[f]{Yuma Aoki}
\author[f]{Yamato Ito}
\author[f]{Kaito Fukuda}
\author[f]{Honoka Kiyama}
\author[o]{Daiki Aoki}
\author[o]{Kaito Fujisawa}
\author[o]{Yasuyuki Shimizu}
\author[o]{Mayu Higuchi}
\author[i]{Masahiro Fukuda}
\author[p]{Natsuki Sakamoto}
\author[j]{Ryuichi Azuma}
\author[e]{Shun Inoue}
\author[o]{Takayoshi Kohmura}
\author[h]{Makoto Yamauchi}
\author[h]{Isamu Hatsukade}
\author[n]{Hironori Matsumoto}
\author[n]{Hirokazu Odaka}
\author[p]{Tsunefumi Mizuno}
\author[a]{Tessei Yoshida}
\author[a]{Yoshitomo Maeda}
\author[a]{Manabu Ishida}
\author[e]{Takeshi G. Tsuru}
\author[q]{Kazutaka Yamaoka}
\author[r]{Takashi Okajima}
\author[r,u]{Takayuki Hayashi}
\author[s]{Junko S. Hiraga}
\author[t]{Masanobu Ozaki}
\author[a,v]{Tadayasu Dotani}
\author[n]{Hiroshi Tsunemi}
\author[n]{Kiyoshi Hayashida}

\affil[a]{Japan Aerospace Exploration Agency, Institute of Space and Astronautical Science, 3-1-1 Yoshino-dai, Chuo-ku, Sagamihara, Kanagawa 252-5210, Japan}
\affil[b]{Faculty of Science and Engineering, Chuo University, 1-13-27 Kasuga, Bunkyo, Tokyo 112-8551, Japan}
\affil[c]{Department of Physics, Faculty of Science, Tokyo University of Science, Kagurazaka, Shinjuku-ku, Tokyo 162-0815, Japan}
\affil[d]{Astronomical Institute, Tohoku University, 6-3 Aramakiazaaoba, Aoba-ku, Sendai, Miyagi 980-8578, Japan}
\affil[e]{Department of Physics, Kyoto University, Kitashirakawa Oiwake-cho,Sakyo-ku, Kyoto, Kyoto 606-8502, Japan}
\affil[f]{Department of Physics, Kindai University, 3-4-1 Kowakae, Higashi-Osaka, Osaka 577-8502, Japan}
\affil[g]{Department of Physics, University of Tokyo, 7-3-1 Hongo, Bunkyo, Tokyo 113-0033, Japan}
\affil[h]{Faculty of Engineering, University of Miyazaki, 1-1 Gakuen Kibanadai Nishi, Miyazaki, Miyazaki 889-2192, Japan}
\affil[i]{College of Science and Engineering, Kanto Gakuinn University, Kanazawa-ku, Yokohama, Kanagawa 236-8501, Japan}
\affil[j]{Department of Physics, Konan University, 8-9-1 Okamoto, Higashinada, Kobe, Hyogo 658-8501, Japan}
\affil[k]{Faculty of Informatics, Tohoku Gakuin University, 3-1 Shimizukoji, Wakabayashi-ku, Sendai, Miyagi 984-8588, Japan}
\affil[l]{Science Education, Faculty of Education, Shizuoka University, Suruga-ku, Shizuoka, Shizuoka 422-8529, Japan}
\affil[m]{Faculty of Education, Nara University of Education, Nara, Nara 630-8528, Japan}
\affil[n]{Department of Earth and Space Science, Osaka University, 1-1 Machikaneyama-cho, Toyonaka, Osaka 560-0043, Japan}
\affil[o]{Department of Physics, Faculty of Science and Technology, Tokyo University of Science, 2641 Yamazaki, Noda, Chiba 270-8510, Japan}
\affil[p]{Hiroshima Astrophysical Science Center, Hiroshima University, 1-3-1 Kagamiyama, Higashi-Hiroshima, Hiroshima 739-8526, Japan}
\affil[q]{Department of Physics, Nagoya University, Chikusa-ku, Nagoya, Aichi 464-8602, Japan}
\affil[r]{NASA’s Goddard Space Flight Center, Greenbelt, MD 20771, USA}
\affil[s]{Department of Physics, Kwansei Gakuin University, 2-2 Gakuen, Sanda, Hyogo 669-1337, Japan}
\affil[t]{Advanced Technology Center, National Astronomical Observatory of Japan, Mitaka, Tokyo 181-8588, Japan}

\affil[u]{Center for Space Science and Technology, University of Maryland, Baltimore County (UMBC), Baltimore, MD 21250, USA}
\affil[v]{Department of Space and Astronautical Science, School of Physical Sciences, SOKENDAI (The Graduate University for Advanced Studies), 3-1-1 Yoshino-dai, Chuou-Ku, Sagamihara, Kanagawa 252-5210, Japan}

\authorinfo{E-mail: suzuki.hiromasa@jaxa.jp}

\begin{document} 
\maketitle

\begin{abstract}
XRISM (X-Ray Imaging and Spectroscopy Mission) is an astronomical satellite with the capability of high-resolution spectroscopy with the X-ray microcalorimeter, Resolve, and wide field-of-view imaging with the CCD camera, Xtend. Xtend consists of the mirror assembly (XMA: X-ray Mirror Assembly) and detector (SXI: Soft X-ray Imager). {The SXI is composed of} CCDs, analog and digital electronics, and a mechanical cooler. After the successful launch on September 6th, 2023 (UT) and subsequent critical operations, the mission instruments were turned on and set up. The CCDs have been kept at the designed operating temperature of $-110$\degc ~after the electronics and cooling system were successfully set up. During the initial operation phase, which continued for more than a month after the critical operations, we verified the observation procedure, stability of the cooling system, all the observation options with different imaging areas and/or timing resolutions, and {time-tagged and automated operations including those for South Atlantic Anomaly passages}. We optimized the operation procedure and observation parameters including the cooler settings, imaging areas for the {small window modes}, and event selection algorithm. We summarize our policy and procedure of the initial operations for the SXI. We also report on a couple of issues we faced during the initial operations and lessons learned from them.

\end{abstract}

\keywords{X-rays, X-ray astronomy, XRISM, Xtend, SXI, X-ray CCD}

\section{Introduction: Soft X-ray Imager (SXI) onboard XRISM}
\label{sec-intro}  % \label{} allows reference to this section

X-Ray Imaging and Spectroscopy Mission (XRISM)\cite{tashiro18, tashiro20, tashiro24} is the Japan’s 7th X-ray astronomical satellite mission, which is equipped with the microcalorimeter ``Resolve''\cite{ishisaki18} and CCD camera ``Xtend''\cite{hayashida18, mori23, mori24} in combination with the X-ray mirror assemblies\cite{hayashi24, tamura24}. XRISM was successfully launched into a low-earth orbit on September 6th, 2023 (UT). The focal plane detector of Xtend is the Soft X-ray Imager (SXI), which {consists of} back-illuminated, fully-depleted X-ray CCDs \cite{tanaka18, nakajima20}. Xtend is designed to cover a large field of view ($38'\times38'$) with an angular resolution of $< 1\farcm7$. The required energy resolution is $< 250$~eV at 6~keV, typical for X-ray CCD sensors.

The SXI is composed of SXI-DE (Digital Electronics), SXI-PE (Pixel processing Electronics), SXI-S (sensor part), SXI-S-1ST (mechanical cooler), and SXI-CD (cooler driver). SXI-S mainly consists of the {analog electronics}, four CCD chips installed on the cold plate, and 1ST (single-stage Stirling cooler).
SXI-PE has two mission I/O (MIO) boards, and each runs two CCDs by generating CCD clocking patterns.
{The analog electronics includes SXI-S-FE (Front-end Electronics), which are composed of four CCD driver boards connected to the four CCDs and generate analog clocks, and the video boards, which are composed of four analog ASICs and field-programmable gate array (FPGA) and digitize the signals output by the CCDs.}
Each CCD chip is divided into two segments ``AB'' and ``CD'', for which frame images are read out individually.
The designed operating temperature of the CCDs is $-110$\degc.
The {incident surface of the} CCDs is {coated} with the optical blocking layer, {a 200-nm-thick layer of aluminum\cite{uchida20}}. {This is twice thicker than that used in Hitomi SXI\cite{tanaka18} and is effectively similar to the medium filter of XMM-Newton EPIC\cite{struder01}.}
% SXI-S has a contamination blocking filter on the top of the hood. This is kept at a high temperature $\sim 25$\degc.
%
{SXI-S-1ST is a single-stage Stirling cooler, which consists of a cold head and compressor. An active balancer, which reduces vibration, is inside the cold head.} 

We have four observation modes of SXI, the full-window, 1/8-window, 0.1-s burst, and 1/8-window+burst modes. The frame cycle is 4~s for all the four modes. The frame exposure is 3.96~s for the full-window mode, and is reduced to 0.46~s, 0.06~s, and 0.06~s for the 1/8-window, 0.1-s burst, and 1/8-window+burst modes, respectively.
The two modes with a small imaging area (1/8-window and 1/8-window+burst) are intended for bright point-like sources, to avoid the pile-up of photons.\cite{yoneyama24}
The 0.1-s burst mode is primarily for calibration purposes.
To achieve the desired energy resolution, we artificially inject charges into the imaging area\cite{uchiyama09, nobukawa14, kanemaru20}. We inject charges into every 160th physical row, which corresponds to 80th logical row after the on-chip $2\times2$ binning.

\section{Initial operation plan}
\subsection{Overview}
We planned to begin operations to start up the SXI on Oct. 17, 2023 (UT), 40 days after the launch, when the satellite bus system would have been started up and the critical operations would have been completed.
Table~\ref{tab-plan} summarizes our initial operation plan.
We planned to turn on the electronics in the first four days, then start cooling of the CCDs. 
% After the temperature control is enabled, 

The verification items during the initial operation phase
% including the effective field of view, energy range, and energy resolution, 
are summarized in Table~\ref{tab-veri}. The verification items corresponding to individual operations or observations are listed in Table~\ref{tab-plan}.
During this phase, we planned to optimize some parameters if related requirements were not satisfied. The candidate optimization items and related requirements are listed in Table~\ref{tab-opt}.
In the following subsections, we describe important operational items in detail.

\begin{table}[htb]
    \centering
    \caption{Summary of the SXI initial operation plan}\label{tab-plan}
    \begin{tabular}{l l l c}
    \hline\hline
        Date (UT) & Operation & Note & Verification \\ 
        & & & item(s) \\ \hline
        2023 Oct. 17 & SXI-PE, SXI-DE on  & & (1), (6) \\ 
        2023 Oct. 18 & SXI-S FE, SXI-S Video on & & (1), (6) \\
        2023 Oct. 18--19 & CCD health check ($\sim 0$\degc) & Full-window mode & (2), (4), (6) \\
        2023 Oct. 20 & SXI-CD on & & (3), (5) \\
        2023 Oct. 21 & Cooling parameter adjustment & & (3) \\
        2023 Oct. 21--22  & CCD health check ($-110$\degc) & Full-window and 0.1-s burst modes & (2) \\
        2023 Oct. 22--24 & Point- or diffuse-source observations & To determine 1/8-window position & (7)  \\
        2023 Nov. 1--2 & CCD health check ($-110$\degc)  & 1/8-window and 1/8-window \\
        & & +burst modes & (2) \\
        2023 Nov. 1 & 1/8-window position adjustment  & 1/8-window mode & (7) \\
        2023 Nov. 2 & 1/8-window position adjustment  & 1/8-window+burst mode & (7) \\
        2023 Nov. 9 & Automated commands for  \\
        & South Atlantic Anomaly  \\ 
        & passages enabled & & (5) \\ 
        Before 2024 Feb. & Point-source observations & For verification of  \\
        && scientific performance & (7), (9), (10), (!2) \\
         & Extended-source observations & For verification of & \\
        && scientific performance & (8), (11) \\
         \hline
    \end{tabular}
\end{table}

\begin{table}[htb]
    \centering
    \caption{Verification items during the SXI initial operation phase}\label{tab-veri}
    \begin{tabular}{c l l l}
    \hline\hline
        No. & Verification item & Requirement & Result \\ \hline
        (1) & Health of electronics & work as designed  & \checkmark \\
        (2) & Health of CCDs & performance consistent with ground tests  & \checkmark \\
        % && performance consistent with ground tests  \\
        (3) & Cooling performance & CCDs must be kept at $\leq -110${\degc}  & \checkmark\\
        (4) & Time-tagged commands & work as designed  & \checkmark$^a$ \\
        (5) & Automated commands & work as designed  & \checkmark$^a$ \\
        (6) & Housekeeping/science data & available as designed  & \checkmark \\
        (7) & 1/8-window position & center of the window along RAWY &  \\
        && must be within optical axis $\pm 13\farcs2$ & \checkmark \\
        (8) & Effective field of view & $> 22' \times 22'$ centered on the optical axis & \checkmark \\
        (9) & Effective energy range & 0.4--13~keV & \checkmark \\
        (10) & Angular resolution & $< 1\farcm7$ (HPD) & \checkmark  \\
        (11) & Energy resolution & $< 250$~eV at 6~keV (FWHM)  & \checkmark \\ 
        (12) & Effective area & $> 300$~cm$^2$ at 1.5~keV, $> 270$~cm$^2$ at 6.0~keV  & \checkmark \\
        (13) & Non X-ray background level & $< 10^{-6}$~keV$^{-1}$~s$^{-1}$ arcmin$^{-2}$ cm$^{-2}$  & \checkmark\\
        \hline
    \end{tabular}
    \\$a$ Operations for South Atlantic Anomaly passages were originally planned to work as automated commands, but actually switched to time-tagged commands due to an issue (see Section~\ref{sec-com}). 
    \vspace{1cm}
\end{table}

\begin{table}[htb]
    \centering
    \caption{Optimization items during the SXI initial operation phase}\label{tab-opt}
    \begin{tabular}{l c}
    \hline\hline
        Optimization item & Related requirement(s) \\ \hline
        Cooler operating voltage & (3) \\
        1/8-window position & (7)  \\
        Event selection/dark calculation parameters & (9), (11), (12)  \\
        \hline
    \end{tabular}
    \end{table}

\subsection{Starting up the electronics}
Here we summarize the procedure of the start-up operations of the electronics, SXI-PE, SXI-DE, SXI-S-FE, and SXI-CD.
After the four FEs are turned on, the CCD temperature monitoring becomes available. Then we turn on SXI-CD and start cooling of the CCDs.
The health check for the electronics involved verification of current and power levels consistent with those in ground tests, and nominal command reception and telemetry output behavior.
In case the electronics would show unexpected behaviors, we prepared contingency operations, i.e., commands to turn off or reset each of the electronics components and to clear the error status.

\subsection{CCD health check and countermeasure against ``anomalous charge'' problem}
We confirmed the health of the CCDs by examining raw frame images. The health check involved confirmation of the telemetry format including the correct time stamps, mean pulse heights and standard deviations that are consistent with those from ground tests, and the absence of unidentified image structures including unidentified pinholes or ``bad'' pixels (the pixels always outputting pseudo events and cannot be used for X-ray detection).
We planned to test the full-window mode before beginning the CCD temperature control, and all the four modes at $-110$\degc.

\begin{figure}[htb!]
    \centering
    \includegraphics[width=12cm]{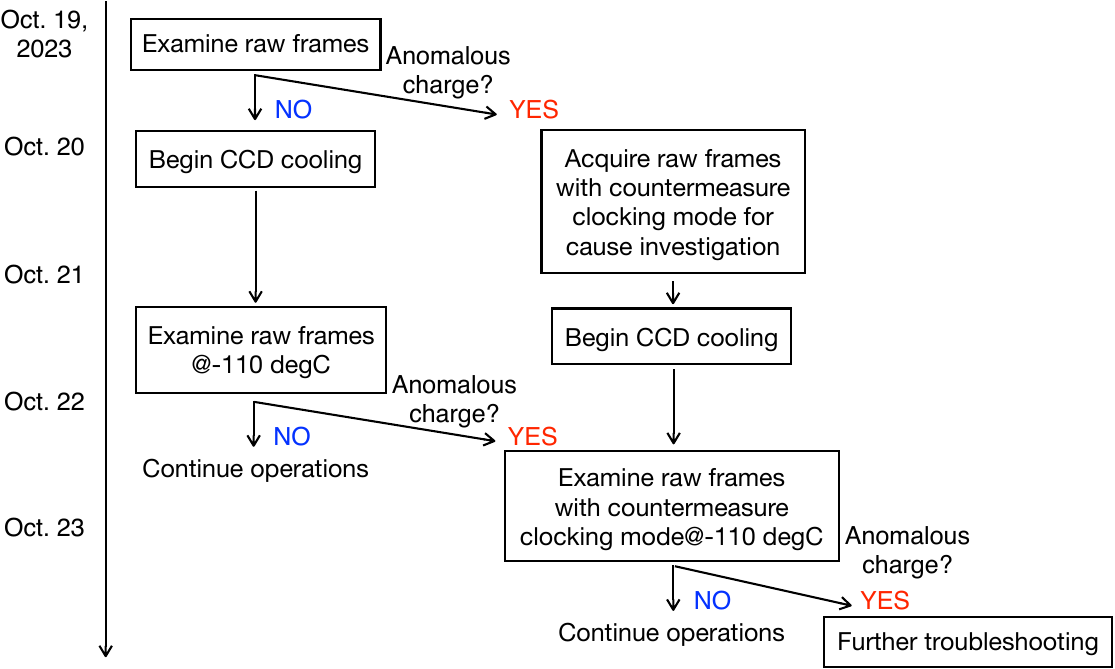}
    \caption{In-orbit operation procedure against the anomalous charge problem. In reality, we did not experience the anomalous charges, and thus we followed the simplest (leftmost) case.}
    \label{fig-anomaly}
\end{figure}

Because we sometimes experienced the ``anomalous charge'' problem\footnote{Large number of charges seem to be injected into the imaging area, making a large fraction of the pixels in a frame saturated with the pulse height of 4095 ch. The cause is yet to be identified. See Noda et al., 2025\cite{noda24jatis} for details.} in on-ground tests, we had to prepare a countermeasure procedure against it in case the problem appears in orbit as well.
Figure~\ref{fig-anomaly} is the flow chart of this procedure. We planned to switch to the countermeasure clocking modes if the anomalous charges appeared in the raw frame images either before or after the temperature control started.
The decision on whether the anomalous charge issue is taking place or not is easy: raw frame images would exhibit some pixels with the pulse height fixed at 4095 ch in frames affected by the anomalous charges.
Other contingency commands related to the CCD health check included those to stop the CCD clocking immediately and to reduce the back-bias voltage.

\subsection{Verification of the performance of CCD temperature control}
After the CCDs were turned on, we planned to confirm the cooling performance of the cooler and its stability. The cooling system must keep the operating temperature of the CCDs, $-110$\degc. In case the cooling performance would not be sufficient, we prepared operation procedures to adjust the cooler operating voltage. 
We also planned to enable automated commands to protect SXI-CD from overcurrent.
The contingency commands included those to disable the automated operation above and to turn off the cooler and/or heater immediately.

\subsection{Adjustment of the 1/8-window position}
Adjustment of the aim point of XRISM was done for the other instrument, Resolve\cite{kanemaru24, maeda24} before we started the operations for the SXI. This slight correction was not important for the SXI given its large field of view in the case of the full-window and 0.1-s burst modes.
On the other hand, it was our task to adjust the window position for the 1/8-window and 1/8-window+burst modes.
In order to sufficiently capture the emission of a point-like source within the 1/8-window region, we had a requirement on the 1/8-window position relative to the optical axis (Table~\ref{tab-veri}). We prepared operation procedures to adjust the window position by changing the CCD clocking to satisfy this requirement.

% Based on the peak position of the diffuse celestial source Abel~2319, observed in Oct. 22--24, 2023, we determined the offset between the peak and central position of the original window region. The offset was found to be 
% $\approx 15$~logical pixels, corresponding to 
% $\approx 28$~arcsec. We then corrected the CCD clocking pattern for the 1/8-window and 1/8-window+burst modes to shift the window position, and tested them with our ground test system composed of EM (Engineering Model) CCDs before using them in orbit.

\subsection{{Verification of the time-tagged/automated commands}}
Nominal operations for the SXI mostly use time-tagged or automated commands.
Time-tagged commands include operations for the day-earth occultation periods, fetching a snapshot of the complete telemetry data from the electronics ($\approx$ once per day), and regular initialization of dark levels ($\approx$ once per day).
The regular initialization of dark levels is needed to reset incorrect dark levels mainly caused by electronic crosstalk between two adjacent segments when a large number of charges is generated by a charged particle\cite{nakajima18}.
% ({sometimes referred to as the} ``cosmic-ray echo'' event\cite{nakajima18}).
Large negative signals resulting from the crosstalk cause errors in the dark-level calculation, resulting in generating persistent, low energy ($\lesssim 1$~keV) events even without true X-ray inputs.
The frequency of this initialization, $\approx$ once per day, was defined as a compromise between operational complexity, data processing requirements, and the need to reduce telemetry size.
The operations for the day-earth occultation include an interruption of the dark-level calculation during the occultation and restarting of the calculation after resuming the nominal observation.
To estimate pixel-dependent and time-variable dark levels, SXI-PE examines every frame and {updates} the stored dark levels if the pulse heights meet certain criteria\cite{tanaka18}.
When Xtend is looking at the day earth, bright optical lights 
% and albedo X-rays 
from the Earth make signals. This could affect the dark-level calculation of SXI-PE. Thus, we stop the dark-level calculation during these periods, {while not interrupting the data acquisition}. 
% {Note that we do not stop data acquisition during these periods.}

Automated commands, which are automatically issued when specified conditions are satisfied, include overcurrent protection commands for the cooler and operations for South Atlantic Anomaly (SAA) passages. The operations for SAA passages include reduction of the back-bias voltage and interruption of the data acquisition by stopping the CCD clocking before entering SAA, and resuming the nominal observation after the passage.
When resuming the nominal observation, SXI-PE first runs a dedicated CCD clocking mode to wipe out stored charges in the imaging area during the SAA passage (CCD ``erasing'' mode) and then resumes the clocking for the nominal observation.
We planned to test if these operations work as designed.

The contingency commands included those to manually reset dark levels when incorrect dark levels are significantly affecting the telemetry, and those to disable the automated commands. The time-tagged commands are uploaded to the spacecraft nominally once per day. Thus, although we can update the time-tagged commands starting from the next uploading ($\lesssim $one day), we cannot modify the commands before that time.

\subsection{{Verification of the scientific performance}}
Scientific performance to be verified during the initial operation phase included the active field of view, energy range, angular resolution, energy resolution, effective area, and non-X-ray background level.
The requirements for these items listed in Table~\ref{tab-veri} were defined to guarantee the sufficient scientific value of the Xtend data.
The active energy range, angular resolution, and effective area were planned to be tested by observing a point-like source. The active field of view can be evaluated by examining an image of an extended source. The spectra of the onboard calibration sources were used to verify the required energy resolution.
For XRISM, night-earth occultation data---typically accumulated to a few Ms per year---can be used to evaluate the non-X-ray background level.
We note that the verification of the quantitative optical blocking performance is outside the scope of the initial operation phase, although the examination of raw frame images would partly cover this verification.

We planned to optimize some of the event-selection and dark-calculation parameters if some of the active energy range, effective area, and energy resolution would not satisfy the requirements.
The candidate parameters to optimize included the split threshold for the event reconstruction, upper and lower bounds for pulse heights to be included in the dark calculation, among others.
In addition, we prepared operation procedures to mask bad pixels/columns
% which are those always outputting pseudo events and cannot be used for X-ray detection
\footnote{{The bad columns are CCD columns (pixel arrays along the readout direction) where most of the pixels are bad pixels.}}
to remove useless event data, which overload the telemetry.

\section{{Initial operation results}}
\subsection{Overview}
We began the operations to start up the SXI on Oct. 17, 2023 (UT) {as planned}. Table~\ref{tab-sum} summarizes our initial operation results.
We turned on the electronics in the first four days, then started cooling of the CCDs.
We spent about a month for optimization and verification of operation and observation parameters.
The SXI has been in the nominal operation phase from Feb. 2024, but we decided to make an additional adjustment of the charge-injection rows in Mar. 2024.
We describe each procedure in the following subsections.

\begin{table}[htb]
    \centering
    \caption{Summary of the SXI initial operation {results}}\label{tab-sum}
    \begin{tabular}{l l l}
    \hline\hline
        Date and Time (UT) & Operation & Note \\ \hline
        2023 Oct. 17 03:08 & SXI-PE on &  \\
        2023 Oct. 17 03:09 & SXI-DE on &  \\
        2023 Oct. 18 04:10 & SXI-S FE on & CCD temperature monitoring started \\
        2023 Oct. 18 05:47 & SXI-S Video on &  \\
        2023 Oct. 18--19 & CCD health check ($\sim 0$\degc) & Full-window mode verified \\
        2023 Oct. 20 03:10 & SXI-CD on & CCD heater control started \\
        2023 Oct. 20 04:08 & CCD cooling started &  \\
    2023 Oct. 21--22  & CCD health check ($-110$\degc) & Full-window and 0.1-s burst modes \\
    & & verified \\
        2023 Oct. 21 23:48 & CCD temperature adjusted & $-111.3${\degc} to $-110.0${\degc} \\ 
        2023 Oct. 22--24 & Observation of Abell 2319 & To determine the 1/8-window position  \\
        2023 Nov. 1--2 & CCD health check ($-110$\degc)  & 1/8-window and 1/8-window+burst \\
        & & modes verified \\
        2023 Nov. 1 21:31 & 1/8-window position adjusted  & 1/8-window mode \\
        2023 Nov. 2 16:29 & 1/8-window position adjusted  & 1/8-window+burst mode \\
        2023 Nov. 2 18:07 to Nov. 9 03:48 & Observation interrupted & See Section~\ref{sec-ll1} \\
        2023 Nov. 9--12 & Automated SAA commands tested & Issue found (See Section~\ref{sec-com}) \\
        2023 Nov. 18 14:50 & Bad columns masked & CCD3, Segment~CD \\
        2023 Dec. 2--4  & Observation of NGC 4151 & To verify the 1/8-window position, etc.  \\
        2024 Mar. 10 06:27 & Charge Injection rows adjusted & See Section~\ref{sec-ci} \\
        2024 Mar. 10 06:26 to Mar. 11 00:51 & Incorrect dark-level calculation & See Section~\ref{sec-ll2} \\
        % 2024 Mar. 18 02:21 to Mar. 18 14:03 & Dark level anomaly & CCD2, segment 1, ACTY=432--434 \\
        
         \hline
    \end{tabular}
\end{table}

\subsection{Starting up the electronics}
\begin{figure}[htb!]
    \centering
    \includegraphics[width=16cm]{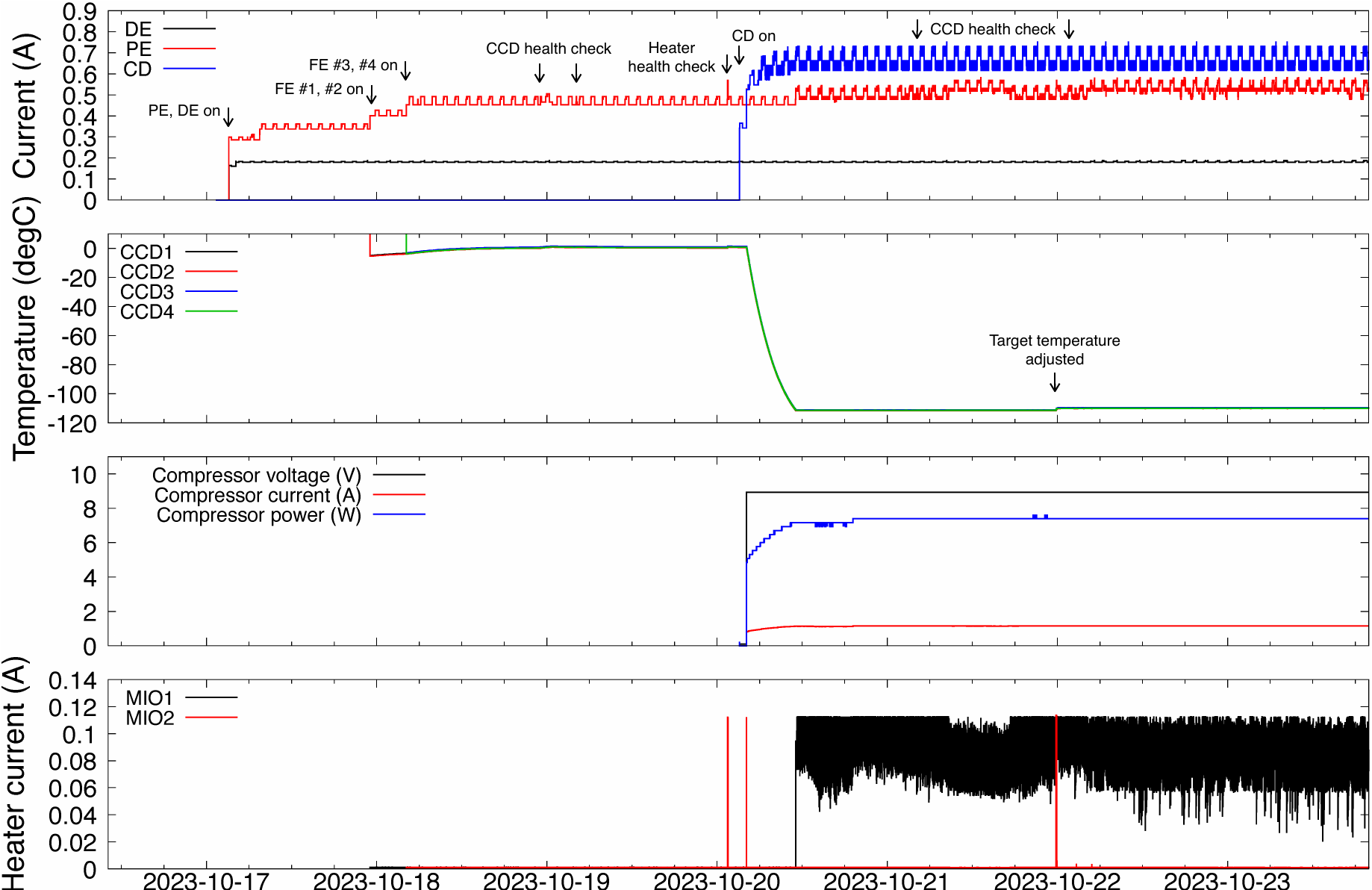}
    \caption{Timeline chart of the SXI house-keeping data related to the electronics and cooling system during the start-up operations. {The vertical axis of the third panel from the top indicates three different quantities as described in the panel.}}
    \label{fig-elec}
\end{figure}

Figure~\ref{fig-elec} shows the timeline chart of our operations starting from the startup of SXI-PE to the temperature stabilization of the CCDs at $\approx -110$\degc.
{We confirmed that the current and power of each component were consistent with the values from on-ground tests when we turned each on.} We note that the $\approx 1.5$~hr variability of the current of the electronics seen in the top panel of Figure~\ref{fig-elec} is due to the variable spacecraft bus voltage, which depends on the satellite location and direction of the Sun.
After the four FEs were turned on, the CCD temperature monitoring became available. Then we turned on the SXI-CD and started cooling of the CCDs.

\subsection{CCD health check and countermeasure against ``anomalous charge'' problem}
\begin{figure}[htb!]
    \centering
    \includegraphics[width=14cm]{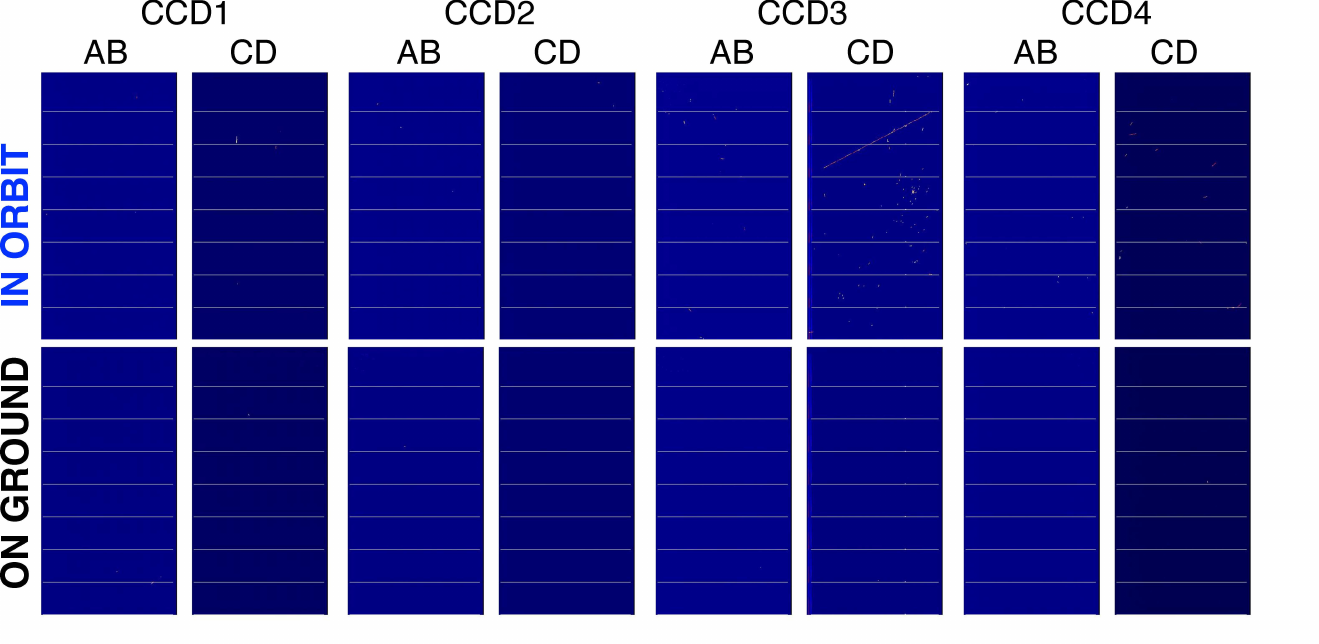}
    \vspace{5pt}
    \caption{In-orbit raw frame images {(upper panels)} at $-110$\degc~ and a comparison with on-ground frames {(lowert panels)}. The eight white rows in each frame are the charge-injection rows, where the pulse heights are saturated at 4095~ch. The color scale is the same for all the frames (0--4095~ch). The white dots and tracks seen in the in-orbit frame images are cosmic-ray induced events.}
    \label{fig-rframe}
\end{figure}

We confirmed the health of the CCDs by examining raw frame images.
We tested the full-window mode before beginning the CCD temperature control, and all the four modes at $-110$\degc.
Figure~\ref{fig-rframe} shows in-orbit raw frame images at $-110$\degc~ obtained on Oct. 21, 2023, which are almost identical to those obtained on ground except for cosmic-ray induced events.
{We found no unidentified image structures including any evidence for the anomalous charges, or increased pinholes or bad pixels.}

\subsection{Verification of the performance of CCD temperature control}
The CCD temperature has been controlled since Oct. 20, 2023 (Figure~\ref{fig-elec}). We confirmed the cooling performance of the 1ST and its stability. In combination with the two heaters attached to the cold plate, the cooling system successfully keeps the operating temperature of the CCDs, $-110$\degc. We note that the original target temperature ($-111.3$\degc) was found to be slightly offset from the desired one, {at which all the ground calibration measurements were conducted}. We thus adjusted it on Oct. 21.
%
% In Figure~\ref{fig-elec}, one can see that only the heater of MIO1 is working mostly. {Since all the CCD chips are on the same cold plate, the heater of MIO1 can manage the temperatures of all the CCDs if the cooling power is well adjusted.}
{The heater of MIO1 maintains the temperatures of all the CCDs, which are on the same cold plate, if the cooler power is appropriate. If more heating power is required to keep the target temperature, the MIO2 heater starts to work.
}
In Figure~\ref{fig-elec}, one can see that only the heater of MIO1 is working mostly,
{indicating that the cooler power is appropriate. Thus, we did not need to adjust the cooler operating voltage.}

\subsection{Adjustment of the 1/8-window position}
\begin{figure}[htb!]
    \centering
    \includegraphics[width=16cm]{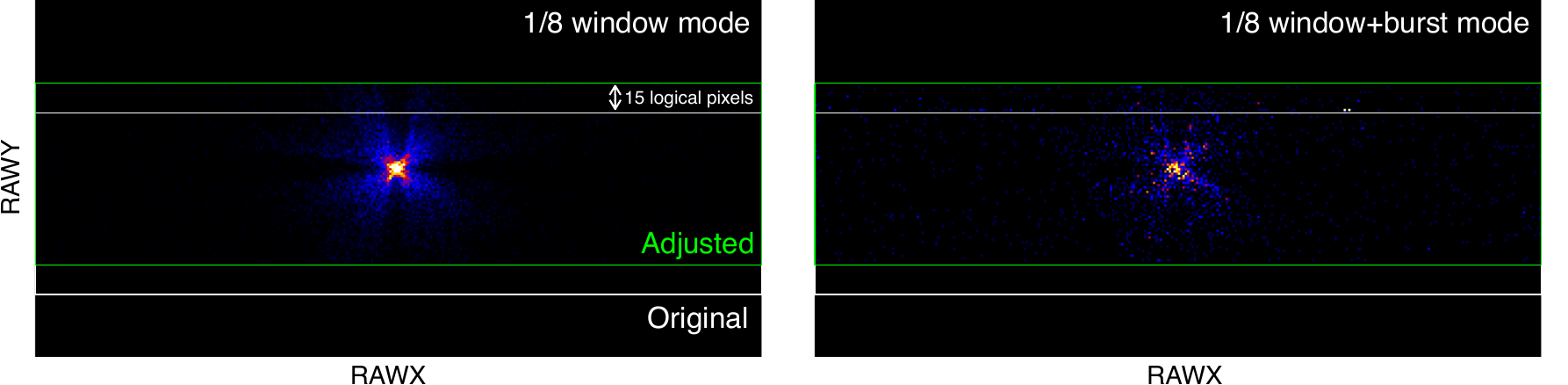}
    \caption{Point-source images obtained with the 1/8-window (left) and 1/8-window+burst (right) modes after the adjustment of the window position. The white and green boxes indicate the window regions before and after the adjustment, respectively.}
    \label{fig-window}
\end{figure}

Based on the peak position of the diffuse celestial source Abell~2319, observed in Oct. 22--24, 2023, we determined the offset between the peak and central position of the original window region. The offset was found to be 
% $\approx 15$~logical pixels, corresponding to 
$\approx 28$~arcsec, {which did not satisfy the requirement (Table~\ref{tab-veri})}. We then corrected the CCD clocking pattern for the 1/8-window and 1/8-window+burst modes to shift the window position, and tested them with our ground test system composed of EM (Engineering Model) CCDs before using them in orbit {starting from Nov. 1 (Table~\ref{tab-sum})}.
The corrected window position was then verified with the observations of the point-like source NGC~4151 in Dec. 2--4, 2023 (Figure~\ref{fig-window}).
{The $\approx$one-month separation between the 1/8-window adjustment and verification was due to the scheduling requirements of the commissioning of other instruments and initial science observations.}

\subsection{{Verification of the time-tagged/automated commands}}\label{sec-com}
\begin{figure}[htb!]
    \centering
    \includegraphics[width=16cm]{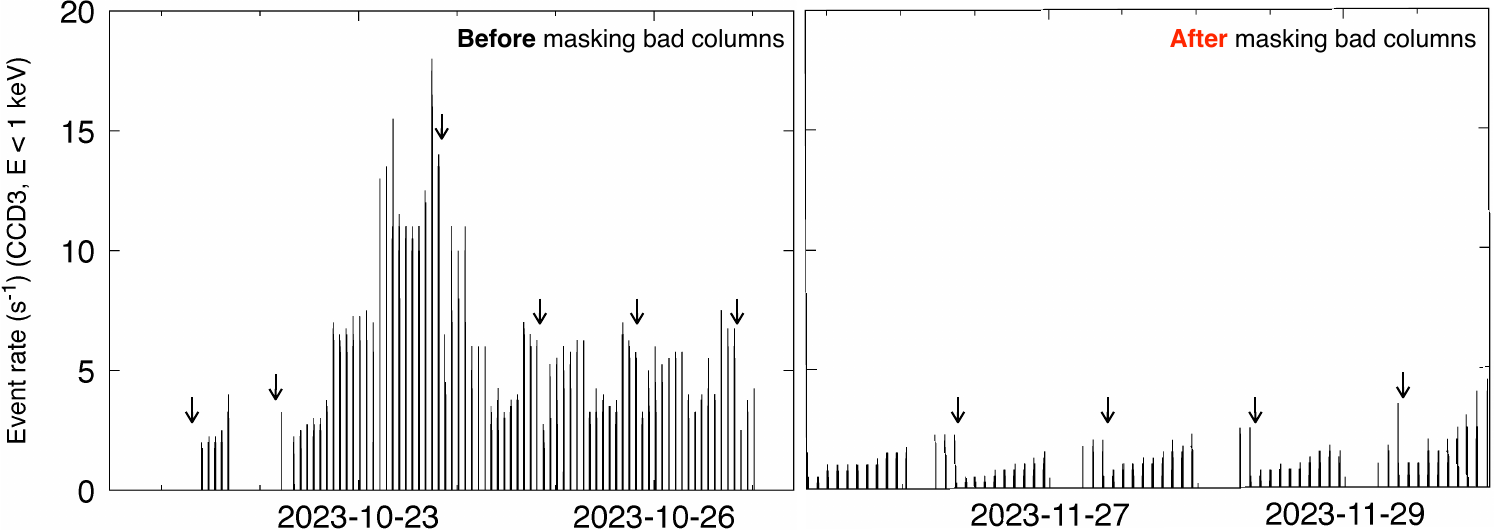}
    \caption{Event rate of CCD3 during night-earth observations before and after masking bad {columns on Nov. 18}. The black arrows indicate time at which dark levels were initialized.}
    \label{fig-initdark}
\end{figure}

{We evaluated whether the time-tagged and automated commands work as designed. All the time-tagged operations, i.e., operations for the day-earth occultation, complete telemetry snapshot, regular initialization of dark levels, were verified.}
{We note that, we experienced an overload of telemetry due to incorrect dark levels before moving on to the nominal operations of the time-tagged commands.}
In Figure~\ref{fig-initdark}, we show the effect of incorrect dark levels and their initialization.
The increasing low energy event rates, most prominent on Oct. 23, 2023, are mostly due to an increasing number of pixels with such incorrect dark levels. Such enhanced pseudo events should be suppressed because they put a load on the data recorder.
When the dark levels are initialized, the event rates are evidently suppressed as expected. Note that the apparent ``offset'' in the event rates in the left panel of Figure~\ref{fig-initdark} is due to bad {pixels}, which persistently generate pseudo events, {and were mostly masked on Nov 18}.

% In order to suppress degradation of the CCDs due to enhanced radiation in the South Atlantic Anomaly (SAA) region and to avoid storing useless data, we stop the CCD clocking and application of the back-bias voltage during SAA passages.
%
{As for the automated operations, the overcurrent protection function for the cooler was successfully enabled.}
Following our original plan, we tried to make {the operations for SAA passages} automatic based on the satellite location {calculated onboard}. During our test of this automated operation starting from Nov. 9, 2023, however, we found that calculated SAA-in and out time was incorrect and thus the operations for {SAA passages} were not working correctly, which later turned out to be due to an {issue in}
% mismatch between the actual orbit parameters and orbit
the orbit calculation algorithm.
Therefore we decided not to use this function based on the satellite location, and rather to issue time-tagged commands.

% When the satellite leaves SAA, SXI-PE first runs a dedicated CCD clocking mode to wipe out stored charges in the imaging area during the SAA passage (CCD ``erasing'' mode) and then resumes the clocking for the nominal observations.

\subsection{{Verification of the scientific performance}}\label{sec-sci}
\begin{figure}[htb!]
    \centering
    \includegraphics[width=14cm]{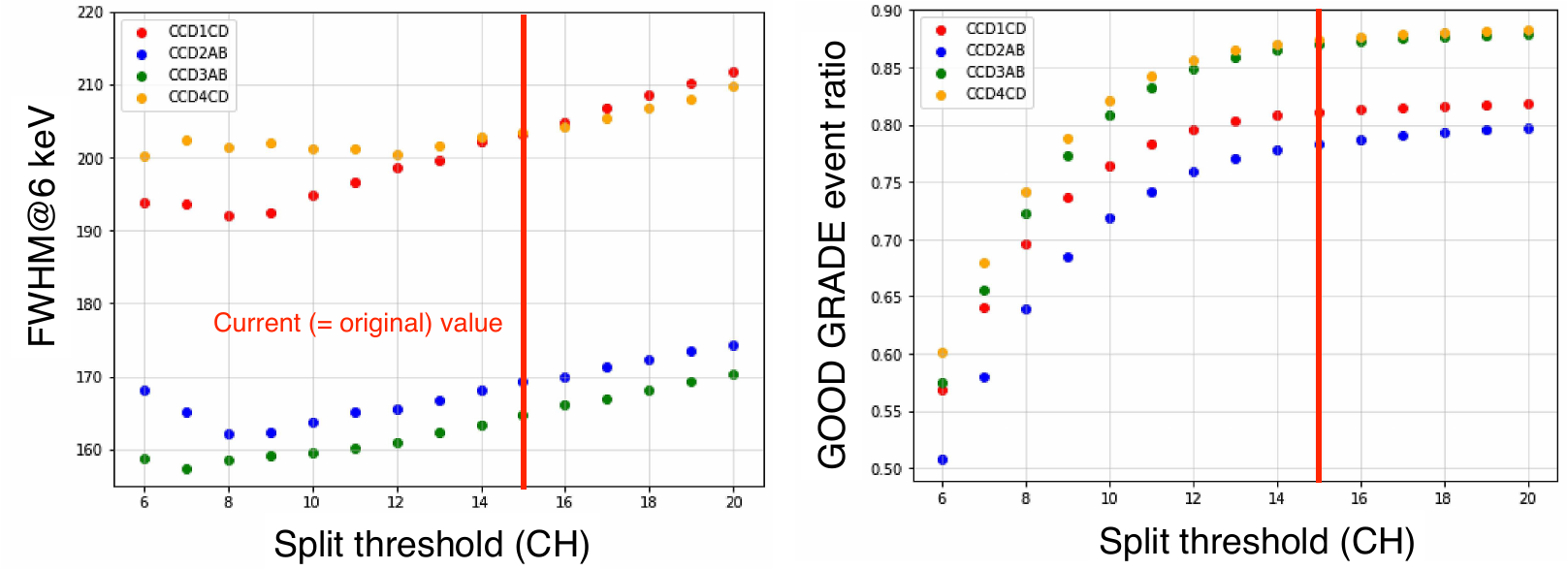}
    \caption{Split-threshold dependence of spectroscopic performance, the energy resolution (left) and good-grade event ratio (right). The red vertical lines are the original value of the split threshold determined before the launch, which was kept unchanged. The energy resolution in the left panel is less than the full performance\cite{mori24} because we used data before the full processing.}
    \label{fig-spth}
\end{figure}

We collected basic imaging and spectral data needed {to evaluate the scientific performance }
% for optimization of the event selection algorithm of all the four observation modes 
by Dec. 2023. Then we scrutinized the data to decide if we have to modify the event selection parameters to optimize the imaging and spectroscopic performance of the SXI.
As a conclusion, we found that {all the requirements are satisfied, and thus nearly all the} parameters were appropriate.
Examination of one of these parameters, the split threshold in the onboard data processing, is as presented in Figure~\ref{fig-spth}. Increasing split threshold values result in degraded spectroscopic performance but more efficient event detection.
The original value turned out to meet the requirements of the energy resolution and effective area.
We made a minor update on the algorithm on the other hand, masking bad columns {in part of the CCD3 Segment CD}, to suppress unimportant events (Figure~\ref{fig-initdark}).
{The scientific performance will be reported separately, in quantitative terms (Uchida et al. in prep.).}

\begin{figure}[htb!]
    \centering
    \includegraphics[width=14cm]{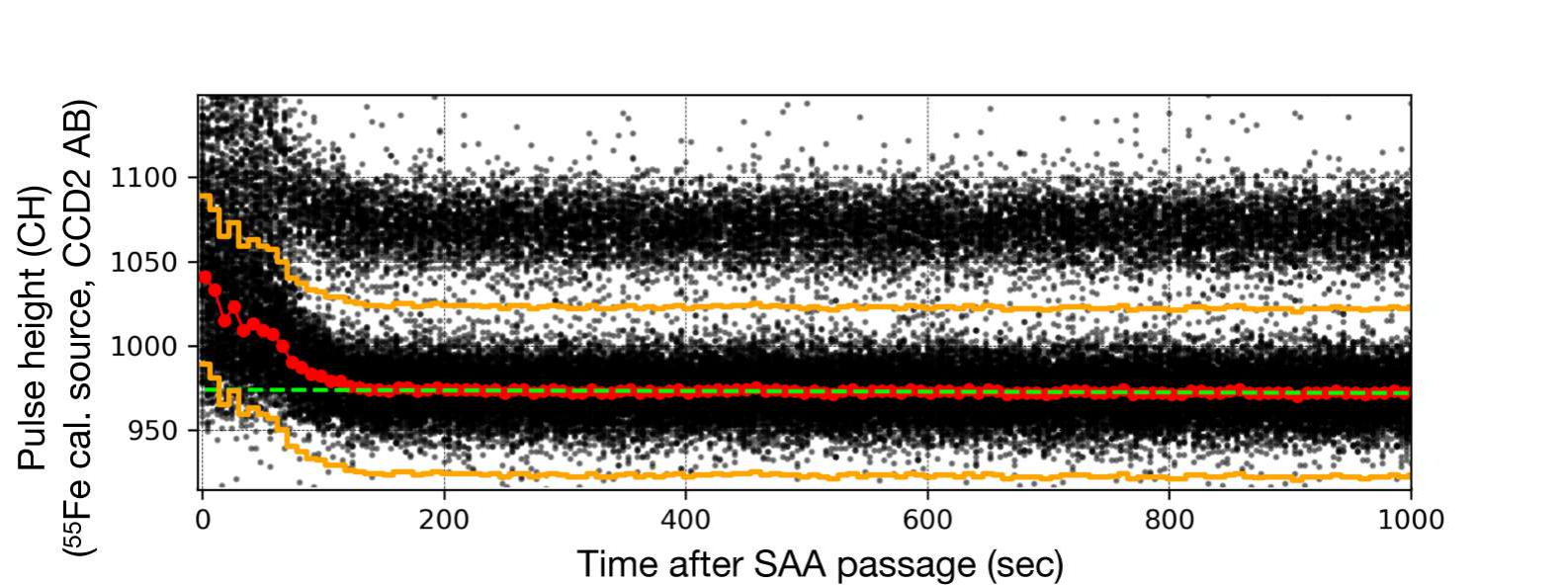}
    \caption{Behavior of pulse heights of Mn-K$\alpha$ and K$\beta$ X-ray events from the $^{55}$Fe calibration source after SAA passages. The black dots are X-ray events. The green dashed line indicates the average pulse height of Mn-K$\alpha$ in the stable period. The Mn-K$\alpha$ centroids at individual time bins, calculated from the data points enclosed with the orange lines, are shown in red.}
    \label{fig-saaout}
\end{figure}

{We found an unexpected behavior of pulse heights after SAA passages.}
Figure~\ref{fig-saaout} shows pulse heights of 5.9~keV and 6.5~keV X-ray events from the onboard $^{55}$Fe calibration source as a function of time after SAA passages. The initial high pulse heights are probably due to the stored charges left even after the CCD erasing, but it was confirmed that the pulse heights become stable after $\lesssim 200$~s. {Although this behavior was unexpected, the time required to resume the nominal performance was found to be smaller than the exclusion time after SAA passages already defined in the standard screening criteria.}

\subsection{Adjustment of the charge-injection rows}\label{sec-ci}
\begin{figure}[htb!]
    \centering
    \includegraphics[width=16cm]{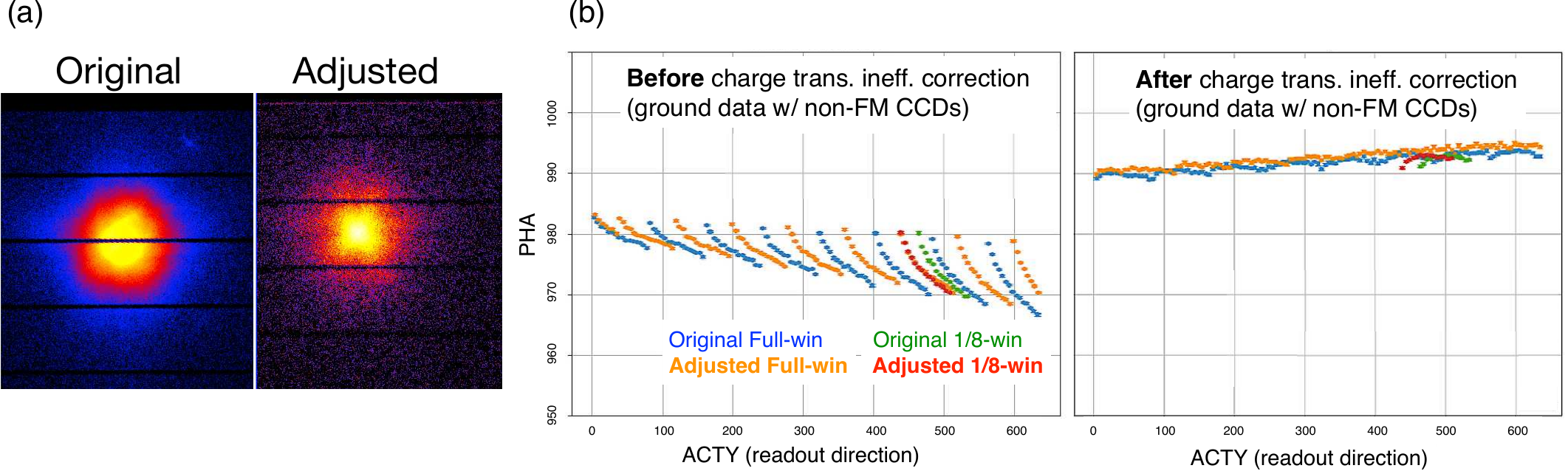}
    \caption{(a) Extended-source images before and after the charge-injection-row adjustment. The black rows without counts are the charge-injection rows. (b) 5.9~keV pulse height distribution along the readout direction before and after correction for charge transfer inefficiency. These are from on-ground experiments to confirm that the original charge-transfer-inefficiency correction algorithm is applicable to the data obtained after the adjustment of the charge injection rows.}
    \label{fig-ci}
\end{figure}

%We basically moved to the nominal operation phase from Dec. 2023. However, there was a strong request to move the charge-injection rows
We officially moved to the nominal operation phase in Feb. 2024. However, we decided to make another special operation after that.
In the full-window mode, one of the charge-injection rows overlaps the aim point (Figure~\ref{fig-ci} (a)). This results in $\sim 10$\% loss of photons for point-like sources.
% Based on the idea that this loss may become unignorable in a long-term point of view, 
Weighing the effort and risk for an additional operation and the scientific merit of it, we decided to adjust the charge-injection rows in Mar. 2024, although this configuration already met all the requirements.

In a similar manner to the adjustment of the window position, we first modified the CCD clocking pattern to shift the charge-injection rows, then tested the modified clocking mode on ground. In this case, we also had to consider potential impact on the charge-transfer-inefficiency correction. So we accumulated data with the ground test system to examine if this adjusted clocking mode had any unexpected impact and if we need to perform additional observations for calibration after performing this adjustment onboard. As a result, we confirmed that the original charge-transfer-inefficiency correction was still applicable to data with the adjusted clocking mode (Figure~\ref{fig-ci} (b)). Thus the spectroscopic performance should be unchanged. Confirmation with in-orbit observations is ongoing and will be reported by Uchida et al in prep.

\section{{Summary of the calibration and performance verification plan}}
Here we briefly summarize our calibration procedure, which is based on the pre-launch plan\cite{miller20}.
The calibration items and relevant data for the SXI are listed in Table~\ref{tab-cal}.
For evaluation of the effective area, we observe point-like sources in coordination with other X-ray telescopes such as NuSTAR.
For the detector response including detector gain and energy resolution, we need line emissions for which exact centroid energies are known. For this purpose, in addition to the onboard $^{55}$Fe calibration sources, we use well-studied supernova remnants and galaxy clusters.
As for the quantum efficiency and contamination, we need to focus on the low energy part, $\lesssim 1$~keV. We also need stable sources, thus low-temperature supernova remnants are one of the most suitable targets. We also use day-earth albedo X-rays, which are expected to be a spatially flat emission, to examine the pixel-to-pixel difference of the contamination.
%We use day-earth observations as well, assuming that the emission spectrum is constant for a few month.
The timing accuracy can be evaluated with the Crab pulsar, for which the pulse period is well known. Note that we use X-rays detected during the frame transfer (out-of-time events) with the 0.1-s burst mode and reconstruct the pulse profile of Crab, thereby we can compare the pulse timing with other instruments (See Nakajima et al., 2018\cite{nakajima18} for details).
As XRISM is a low-earth-orbit satellite, plenty of day- and night-earth observation data are available. The night-earth data represent pure non-X-ray background. The bright and uniform optical lights during day-earth observations can be used for evaluation of the optical blocking performance.

We already obtained basic data sets for all the calibration and performance verification items, and data analysis is ongoing. The resultant performance will be presented by Uchida et al. in prep.

\begin{table}[htb!]
    \centering
    \caption{Calibration items and relevant data for the SXI}
    \begin{tabular}{l l}
    \hline\hline
    Calibration item & Data \\ \hline
    Effective area & 3C273, G21.5$-$0.9 \\
    Detector response & 1E0102.2$-$7219, Perseus, N132D, \\
    & Cygnus Loop, $^{55}$Fe calibration sources \\
    Quantum efficiency & 1E0102.2$-$7219 \\
    Contamination & 1E0102.2$-$7219, Day-earth data \\
    Timing accuracy & Crab  \\
    Non-X-ray background level & Night-earth data \\
    Optical blocking performance & Day-earth data \\ 
    \hline
    \end{tabular}
    \label{tab-cal}
\end{table}

\section{Lessons learned}\label{sec-ll}
\subsection{Need for tests with a ``test as you fly'' environment}\label{sec-ll1}
When we protect the CCDs against SAA passages, we send time-tagged commands to stop the CCD clocking, wait for a few seconds, and then stop the application of the back-bias voltage.
At the time we first tested the 1/8-window mode on Nov. 1--2, 2023, commands to decrease the back-bias voltage were sometimes not accepted because the waiting time was not enough and the clocking was still running.
We had tested these operations a lot of times on ground, but only a few times with the ``test as you fly'' setup, with the same series of the commands as the in-orbit operations.
We had to interrupt observations until we resolved the cause and took measures (Table~\ref{tab-sum}).
After {reviewing the electronics design and past experiments}, we confirmed that the time needed to stop the CCD clocking may sometimes exceed the original wait time.
Therefore, we extended the wait time and resumed observations on Nov. 9.
{Based on this experience, we made our operation preparation procedure more strict. Originally, we verified each of the subsets of commands (i.e., command sequences\footnote{{The command sequences include one to set up the full-window mode and start an observation, one to issue commands required before/after SAA passages, among others.}}) using a ground test system before using them in orbit. After this issue we decided to also verify combinations of the command sequences planned to be used in orbit, such as the consecutive SAA-in and SAA-out operations.}

\subsection{Dark-level calculation}\label{sec-ll2}
Keeping appropriate dark levels is quite important because wrong values lead to a large number of pseudo events as we described in Section~\ref{sec-sci}.
Thus, in our case, regular initialization of dark levels is essential, thereby we reset the {crosstalk} effect.
The high event rates seen on Oct. 23, 2023 in Figure~\ref{fig-initdark} is due to an inappropriate operation where we failed to initialize dark levels once a day.
The other lesson we learned about dark levels is that we should be careful about the status of the dark-level calculation when we conduct specialized operations. On Mar. 10, 2024, when we adjusted the charge-injection rows, we ran the CCD erasing with the dark-level calculation enabled. This resulted in anomalous dark levels in Mar. 10--11 (Table~\ref{tab-sum}). 
{To apply this lesson learned, we improved our operation preparation procedure when planning a specialized operation, to ensure disabling the time-tagged and automated commands which may interfere with the desired operation.}

\section{Summary}
In this paper, we summarized the initial operation procedure of SXI, the focal plane detector of Xtend onboard XRISM. We started our operation on Oct. 17, 2023, 40 days after the launch. The CCDs have been kept at the designed operating temperature of $-110$\degc ~after the electronics and cooling system were successfully set up. We verified the observation procedure, stability of the cooling system, all the observation options with different imaging areas and/or timing resolutions, and {time-tagged and automated operations including those for} SAA passages. We optimized the operation procedure and observation parameters including the cooler settings, window position for the small window modes, and event selection algorithm.
We started observations for calibration and performance verification, the results of which will be presented by Uchida et al. in prep.

\section*{Data availability}
The data that support this work are proprietary and are not publicly available. The data plotted in the figures are available from the corresponding author upon request, and a limited subset of the underlying data can be requested from the author at hiromasa050701@gmail.com.

\section*{Disclosure}
The authors declare there are no financial interests, commercial affiliations, or other potential conflicts of interest that have influenced the objectivity of this research or the writing of this paper.

\acknowledgments % equivalent to \section*{ACKNOWLEDGMENTS}       
This work is based on our SPIE proceedings paper\cite{suzuki24}.
This work was supported by JSPS/MEXT KANENHI Nos. 21J00031, 22KJ3059, 24K17093, 19K21884, 20H01947, 20KK0071, 23K20239, 24K17105, 21H01095, 23K20850, 21K20372, 23K22536.
This work was also supported by JSPS Core-to-Core Program (grant number: JPJSCCA20220002).

% References
\bibliography{report} % bibliography data in report.bib
\bibliographystyle{spiebib} % makes bibtex use spiebib.bst

\end{document}